\documentclass[aps,prl,unsortedaddress,superscriptaddress,a4paper,nofootinbib,nobalancelastpage,showpacs,twocolumn]{revtex4-1}
\usepackage{latexsym,amsbsy}
\usepackage{multirow}
\usepackage[dvips]{graphicx}
\begin{document} 

\title{Strangeness -2 hypertriton} 

\author{H.~Garcilazo} 
\affiliation{Escuela Superior de F\' \i sica y Matem\'aticas, \\ 
Instituto Polit\'ecnico Nacional, Edificio 9, 
07738 M\'exico D.F., Mexico} 

\author{A.~Valcarce}  
\affiliation{Departamento de F\'\i sica Fundamental, \\
Universidad de Salamanca, E-37008 Salamanca, Spain}

\date{\today} 

\begin{abstract} 
We have solved for the first time the Faddeev equations for 
the bound state problem of the coupled 
$\Lambda\Lambda N$--$\Xi NN$ system to study whether
an hypertriton with strangeness $-2$ may exist or not. 
We make use of the interactions obtained from a chiral quark model
describing the low-energy observables of the two-baryon
systems with strangeness $0$, $-1$, and $-2$ and three-baryon systems
with strangeness $0$ and $-1$. 
The $\Lambda\Lambda N$ system alone is unbound. However, when the full coupling to
$\Xi NN$ is considered, the strangeness $-2$ three-baryon 
system with quantum numbers $(I,J^P)=(\frac{1}{2},\frac{1}{2}^+)$ becomes
bound, with a binding energy of about 0.5 MeV. This result is compatible with 
the non-existence of a stable $^3_\Lambda$H with isospin one.
\end{abstract} 

\pacs{21.45.-v, 21.80.+a, 21.10.Dr, 12.39.Jh} 

\keywords{nucleon-hyperon interactions, hyperon-hyperon interactions, 
Faddeev equations} 

\maketitle 

The strangeness $\hat{S}=-2$ sector has become an important issue for 
theoretical and experimental studies of the strangeness nuclear 
physics. The $\Xi N -\Lambda \Lambda$ interaction accounts
for the existence of doubly strange hypernuclei, which is a
gateway to strange hadronic matter. 
The $(K^-,K^+)$ reaction is one of the most promising ways of studying
doubly strange systems. $\Lambda\Lambda$ hypernuclei can be produced
through the reaction $K^- p \to K^+ \Xi^-$ followed by $\Xi^- p \to \Lambda\Lambda$.
Strangeness $-2$ baryon-baryon 
interactions also account for a possible six-quark H dibaryon, 
which has yet to be experimentally observed. The future E07 experiment 
from J--PARC~\cite{Nak10a,Nak10b} is expected 
to improve our knowledge on the $\hat{S}=$--2 sector, giving ten times 
more emulsions events for double$-\Lambda$ hypernuclei. 

On the experimental side, there are very few data in the $\hat{S}=-2$ sector 
coming from the inelastic $\Xi^- p \to \Lambda\Lambda$ cross section at a 
lab momentum of around 500 MeV/c, and from the elastic $\Xi^- p \to \Xi^- p$ 
and inelastic $\Xi^- p \to \Xi^0 n$ cross sections for lab momenta in the 
range of 500 -- 600 MeV/c~\cite{Ahn06,Tam01,Yam01}. 
The relevant information we have is indirect and comes from 
double--$\Lambda$ hypernuclei. Their binding energies, $B_{\Lambda\Lambda}$, 
provide upper limits for that of the H dibaryon, i.e., $B_H < B_{\Lambda\Lambda}$. 
The first hypernuclear events are quite old and admit several 
interpretations~\cite{Dan63,Pro66,Aok91}. 
In 2001 it was reported the so--called Nagara event~\cite{Tak01}, 
interpreted uniquely as the sequential decay of ${}^6_{\Lambda\Lambda}$He 
emitted from a $\Xi^-$--hyperon nuclear capture at rest. The mass and the 
values of $B_{\Lambda\Lambda}$ and of the $\Lambda\Lambda$ interaction 
energy, $\Delta B_{\Lambda\Lambda}$, were determined without ambiguities. 
It gave the most stringent 
constraint to the mass of the H dibaryon to date, i.e., $M_H > 2223.7$ MeV 
at a 90\% confidence level. It took almost one decade, but four more 
double--$\Lambda$ hypernuclear events were reported, from KEK E176 and 
E373 experiments~\cite{Nak10a}, still with preliminary results. 
All the details are summarized in Table~\ref{t0}.
\begin{table}[b]\caption{Double $\Lambda$ hypernuclear events.}
\begin{center}
\begin{tabular}{c  c c c}
\hline\hline
Event & Nuclide & $B_{\Lambda\Lambda}$ (MeV) & $\Delta B_{\Lambda\Lambda}$ (MeV) \\
\hline
1963          & ${}_{\Lambda\Lambda}^{10}$Be & 17.7 $\pm$ 0.4  & 4.3  $\pm$ 0.4   \\
1966          & ${}_{\Lambda\Lambda}^6$He    & 10.9 $\pm$ 0.5  & 4.7  $\pm$ 1.0   \\
1991          & ${}_{\Lambda\Lambda}^{13}$B  & 27.5 $\pm$ 0.7  & 4.8  $\pm$ 0.7   \\
NAGARA        & ${}_{\Lambda\Lambda}^6$He    & 7.13 $\pm$ 0.87 & 1.0  $\pm$ 0.2   \\
MIKAGE        & ${}_{\Lambda\Lambda}^6$He    & 10.06 $\pm$ 1.72& 3.82 $\pm$ 1.72  \\
DEMACHIYANAGI & ${}_{\Lambda\Lambda}^{10}$Be & 11.90 $\pm$ 0.13& --1.52 $\pm$ 0.15 \\
HIDA          & ${}_{\Lambda\Lambda}^{11}$Be & 20.49 $\pm$ 1.15&  2.27 $\pm$ 1.23 \\
              & ${}_{\Lambda\Lambda}^{12}$Be & 22.23 $\pm$ 1.15&   --             \\
E176          & ${}_{\Lambda\Lambda}^{13}$Be &  23.3 $\pm$ 0.7 & 0.6 $\pm$ 0.8 \\
\hline\hline
\end{tabular}
\end{center}
\label{t0}
\end{table}

Besides the double-$\Lambda$ hypernuclei quoted in Table~\ref{t0}, there is a 
general consensus that the mirror $\Lambda\Lambda$ hypernuclei 
${}_{\Lambda\Lambda}^5$H--${}_{\Lambda\Lambda}^5$He are particle stable~\cite{Nem05}.
The existence of a ${}_{\Lambda\Lambda}^4$H bound state has been claimed by the AGS-E906
experiment~\cite{Ahn01}, from
correlated weak-decay pions emitted sequentially by $\Lambda\Lambda$ hypernuclei produced in a $(K^-,K^+)$
reaction on $^9$Be. The stability of the $\Lambda\Lambda N$ systems was discarded long ago~\cite{Tan65}
by using symmetry considerations with respect to the $^3_\Lambda$H, and 
therefore without considering the important coupled channel effect due to the
existence of the $\Xi NN$ system.

Theoretically, the $\hat{S}=-2$ sector was recently put back on the agenda by lattice QCD calculations 
of different collaborations, NPLQCD~\cite{Bea10} and HAL QCD~\cite{Ino10}, 
providing evidence for a $\Lambda\Lambda$ bound state for non--physical values of the 
pion mass ($m_{\pi}$ = 389 MeV and $m_\pi$ = 673 $\to$ 1010 MeV, respectively). 
When performing quadratic and linear extrapolations to the physical point~\cite{Bea11}, 
a bound dibaryon (around 7 MeV) and a H at threshold, 
respectively, are predicted. Ref.~\cite{Bea11} presents preliminary results 
for $m_{\pi}$ = 230 MeV, much closer to the physical pion mass, pointing to a H dibaryon at threshold, 
as also experimentally suggested by the enhancement of the $\Lambda\Lambda$ production near threshold 
found in Ref.~\cite{Yoo07}. 

The purpose of this letter is twofold. On the one hand we present the solution of the
Faddeev equations for the bound state problem of the coupled $\Lambda\Lambda N$--$\Xi NN$ system. 
The system has been formally studied and its 
Faddeev equations written down~\cite{Miy97,Miy01}, 
although they have never been applied in a 
numerical calculation with realistic two-body interactions. This is 
basically due to the fact that one requires a model of the baryon-baryon 
interaction which should be able to simultaneously describe two-baryon 
states with strangeness $0$, $-1$, and $-2$ within a single consistent 
theoretical framework. Afterwards,
we will apply the formalism by means of the interactions obtained from a chiral quark model
describing the low-energy observables of the two-baryon
systems with strangeness $0$, $-1$, and $-2$ and also three-baryon systems
with strangeness $0$ and $-1$, trying to elucidate the nature of the three-baryon
system with strangeness $-2$.

The coupled $\Lambda\Lambda N$--$\Xi NN$ system is peculiar because
it has two identical particles in each of its two components although 
they are of different type, which complicates considerably its analysis.
The Faddeev equations for the bound-state problem of the coupled
$\Lambda\Lambda N$--$\Xi NN$ system have been derived in Ref.~\cite{Miy01}.
We have obtained these same equations by an independent method~\cite{Gar12}, they read:
\begin{widetext}
\begin{eqnarray}
T^{\Xi NN} &=& t^{NN,NN}(1-P_{23})P_{13}P_{23}
G_0^{N\Xi N} T^{N\Xi N},
\nonumber \\
T^{N\Xi N} &=& t^{N\Xi,N\Xi}P_{12}P_{23}
 G_0^{\Xi NN} T^{\Xi NN}
- t^{N\Xi,N\Xi}P_{13}
G_0^{N\Xi N} T^{N\Xi N} +
t^{N\Xi,\Lambda\Lambda}
(1-P_{23})P_{13}P_{23}
 G_0^{\Lambda N\Lambda} T^{\Lambda N\Lambda},
\nonumber \\
T^{N\Lambda\Lambda} &=& t^{\Lambda\Lambda,N\Xi}
P_{12}P_{23}
 G_0^{\Xi NN} T^{\Xi NN}
- t^{\Lambda\Lambda,N\Xi}P_{13}
G_0^{N\Xi N} T^{N\Xi N} +
 t^{\Lambda\Lambda,\Lambda\Lambda}
(1-P_{23})P_{13}P_{23}
 G_0^{\Lambda N\Lambda} T^{\Lambda N\Lambda},
\nonumber \\
T^{\Lambda N\Lambda} &=& t^{N\Lambda,N\Lambda}
P_{12}P_{23}
 G_0^{N\Lambda\Lambda} T^{N\Lambda\Lambda}-
t^{N\Lambda,N\Lambda}P_{13}
 G_0^{\Lambda N\Lambda} T^{\Lambda N\Lambda},
\label{eq4} 
\end{eqnarray}
\end{widetext}
where $G_0^{ijk}$ is the propagator for three free
particles $ijk$, $t^{ij,kl}$ are the two-body $t-$matrices for the different
transitions $ij \to kl$, and $P_{ij}$ is the exchange operator for particles 
$i$ and $j$. The first superscript in the $T-$functions is the spectator
and the other two are the interacting pair.
We will solve these equations including all the $S-$wave
configurations $\ell_i=\lambda_i=0$, where $\ell_i$ is the
orbital angular momentum between particles $j$ and $k$, and
$\lambda_i$ is the orbital angular momentum between 
particle $i$ and the pair $jk$. Therefore, the total
angular momentum $J=1/2$ is equal to the total spin. 

The set of Eqs.~(\ref{eq4}) are integral equations
in two continuous variables which couple the nine two-body 
channels obtained from Table~\ref{t1}. 
In order to solve these equations we use the method
applied in our previous works~\cite{Gar07a,Gar07b},
where the two-body $t-$matrices are
expanded in terms of Legendre polynomials leading to integral
equations in only one continuous variable coupling the various Legendre
components required for convergence.
\begin{table}
\caption{S-wave two-body channels $(i,j)$ of the various 
subsystems that contribute to the strangeness $-2$
$(I,J^P)=(\frac{1}{2},\frac{1}{2}^+)$ three-body state.} 
\begin{ruledtabular} 
\begin{tabular}{cccc} 
& Subsystem & $(i,j)$ channels  & \\
\hline
& $NN$  & (0,1),(1,0)   & \\
& $N\Lambda$  & $(\frac{1}{2},0)$,$(\frac{1}{2},1)$   & \\
& $\Lambda\Lambda$  & (0,0)   & \\
& $N\Xi$  & (0,0),(0,1),(1,0),(1,1)   & \\
\end{tabular}
\end{ruledtabular} 
\label{t1}
\end{table}

In each of the two components of the coupled
$\Lambda\Lambda N$--$\Xi NN$ system we take particles 2 and 3
to be the two identical ones and particle 1 to be the 
different one. We will take the basis states
1 and 3 using a cyclic coupling scheme, i.e., $1=(2+3)+1$, and
$3=(1+2)+3$, while for the basis state 2 we use the anticyclic 
scheme $2=(1+3)+2$. With these conventions, Eqs.~(\ref{eq4}) take 
the explicit form~\cite{Gar12},
\begin{widetext}
\begin{eqnarray}
T^{\Xi NN}_{\alpha_1m}(q_1) &=& 2\sum_{\alpha_3n}\int_0^\infty
q_3^2dq_3
K^{NN,NN; N\Xi N}_{mn;\alpha_1\alpha_3;13}(q_1q_3) 
T^{N\Xi N}_{\alpha_3n}(q_3),
\nonumber \\
T^{N\Xi N}_{\alpha_3m}(q_3) &=& \sum_{\alpha_1n}\int_0^\infty
q_1^2dq_1
K^{N\Xi,N\Xi; \Xi NN}_{mn;\alpha_3\alpha_1;31}(q_3q_1) 
T^{\Xi NN}_{\alpha_1n}(q_1) -
\sum_{\alpha_3^\prime n}\int_0^\infty
{q_3^\prime}^2dq_3^\prime
K^{N\Xi,N\Xi; N\Xi N}_{mn;\alpha_3\alpha_3^\prime;23}(q_3q_3^\prime) 
T^{N\Xi N}_{\alpha_3^\prime n}(q_3^\prime)
\nonumber \\ &+&
2\sum_{\alpha_3^\prime n}\int_0^\infty
{q_3^\prime}^2dq_3^\prime
K^{N\Xi,\Lambda\Lambda; \Lambda N\Lambda}_{mn;\alpha_3\alpha_3^\prime;13}(q_3q_3^\prime) 
T^{\Lambda N\Lambda}_{\alpha_3^\prime n}(q_3^\prime),
\nonumber \\
T^{N\Lambda\Lambda}_{\alpha_1m}(q_1) &=& \sum_{\alpha_1^\prime n}\int_0^\infty
{q_1^\prime}^2dq_1^\prime
K^{\Lambda\Lambda,N\Xi; \Xi NN}_{mn;\alpha_1\alpha_1^\prime;31}(q_1q_1^\prime) 
T^{\Xi NN}_{\alpha_1^\prime n}(q_1^\prime) -
\sum_{\alpha_3 n}\int_0^\infty
q_3^2dq_3
K^{\Lambda\Lambda,N\Xi; N\Xi N}_{mn;\alpha_1\alpha_3;23}(q_1q_3) 
T^{N\Xi N}_{\alpha_3 n}(q_3)
\nonumber \\ &+&
2\sum_{\alpha_3 n}\int_0^\infty
q_3^2dq_3
K^{\Lambda\Lambda,\Lambda\Lambda; \Lambda N\Lambda}_{mn;\alpha_1\alpha_3;13}(q_1q_3) 
T^{\Lambda N\Lambda}_{\alpha_3 n}(q_3),
\nonumber \\
T^{\Lambda N\Lambda}_{\alpha_3m}(q_3) &=& \sum_{\alpha_1 n}\int_0^\infty
q_1^2dq_1
K^{N \Lambda,N\Lambda; N\Lambda\Lambda}_{mn;\alpha_3\alpha_1;31}(q_3q_1) 
T^{N\Lambda\Lambda}_{\alpha_1 n}(q_1) -
\sum_{\alpha_3^\prime n}\int_0^\infty
{q_3^\prime}^2dq_3^\prime
K^{N\Lambda,N\Lambda; \Lambda N\Lambda}_{mn;\alpha_3\alpha_3^\prime;23}(q_3q_3^\prime) 
T^{\Lambda N\Lambda}_{\alpha_3^\prime n}(q_3^\prime),
\label{eq8} 
\end{eqnarray}
where
\begin{eqnarray}
K^{\beta\gamma,\delta\zeta;\kappa\lambda\rho}_{mn;\alpha_i\alpha_j;kl}(q_iq_j)&=&\frac{2m+1}{4}A^{\alpha_i\alpha_j}_{kl}
\int_{-1}^1 d{\rm cos}\theta\int_{-1}^1 dx_i P_m(x_i) P_n(x_j)
\nonumber \\ &\times&
t^{\beta\gamma,\delta\zeta}_{\alpha_i}(p_i,p_i^\prime;E+\Delta E
-q_i^2/2\nu_i)\frac{1}{E+\Delta E-{p_j^\prime}^2/2\eta_j
-q_j^2/2\nu_j}.
\label{eq9} 
\end{eqnarray}
\end{widetext}
$\eta_j$ and $\nu_j$ are the usual reduced masses and
$P_n(x)$ is a Legendre polynomial.
$p_i=b(1+x_i)/(1-x_i)$,
$x_j=(p_j-b)/(p_j+b)$,
and $b$ is a scale parameter on which the solution does not
depend.
$p_i^\prime=[q_j^2+(\eta_iq_i/m_k)^2+2(\eta_iq_iq_j/m_k){\rm cos}
\theta]^{1/2}$, 
$p_j=[q_i^2+(\eta_jq_j/m_k)^2+2(\eta_jq_iq_j/m_k){\rm cos}
\theta]^{1/2}$,
and $\Delta E=0$ if the
corresponding state (either $i$ or $j$) belongs to 
the $N\Lambda\Lambda$
component, while $\Delta E=2m_\Lambda-m_N-m_\Xi$ if the
corresponding state belongs to the $\Xi NN$ component.
Finally, $A^{\alpha_i\alpha_j}_{kl}$ are the usual spin-isospin
transition coefficients~\cite{Gar07a}, where $\delta\zeta$ is the
interacting pair in the state $i$ and $\lambda\rho$ is the
interacting pair in the state $j$.

For practical purposes, we took into account all the $S-$wave two-body amplitudes that contribute
in Eqs.~(\ref{eq8}) as shown in Table~\ref{t1}.
Even though our calculation will include only two-body $S-$waves, the corresponding two-body amplitudes
will be obtained from a full model, including $D$ waves
in spin-triplet channels and the coupling to higher 
mass states in those cases where the quantum numbers
allow for it.

Once the method to solve the bound state problem of the $\Lambda\Lambda N$--$\Xi NN$ system has been
designed, we apply it to the chiral quark model of the baryon-baryon interaction developed
in Ref.~\cite{Val05}. The model is capable to describe the low-energy parameters
of the two-nucleon system, the $S-$wave phase shifts, and the triton binding energy~\cite{Jul02}.
It reproduces the elastic and inelastic scattering cross sections of the $\hat{S}=-1$
two-baryon systems and the hypertriton binding energy~\cite{Gar07a,Gar07b}. As can
be seen in Fig. 2 of Ref.~\cite{Gar07b}, the isospin one $\Lambda NN$ system is unbound.
Finally, the model provides parameter free predictions for the elastic and inelastic 
scattering cross sections of the $\hat{S}=-2$ two-baryon systems~\cite{Val10} that
are consistent with the scarce available data. In particular, the relevant
$\Xi^-p \to \Lambda\Lambda$ is correctly described (see Fig. 2 of Ref.~\cite{Val10}).
Thus, we are confident that the interactions are realistic enough to 
allow for the study of the existence (or non-existence)
of the strangeness $-2$ hypertriton.

The H dibaryon has strangeness $-2$, positive parity, and 
isospin and spin $(i,j)=(0,0)$. It appears in our model as a bound state
of the coupled $\Lambda\Lambda$--$N\Xi$--$\Sigma\Sigma$ system
with a binding energy of 6.928 MeV~\cite{Val10}.
Therefore, the main configuration of the strangeness $-2$ hypertriton
will be an H dibaryon as the interacting pair and a $S-$wave nucleon as 
spectator, which leads to total isospin and spin $(I,J)=(\frac{1}{2},
\frac{1}{2})$ and positive parity. This configuration is also favored 
by having a deuteron as interacting pair and a $S-$wave $\Xi$ hyperon as spectator.
We give in Table~\ref{t1} all the $S-$wave two-body
channels that contribute to the 
$(I,J^P)=(\frac{1}{2},\frac{1}{2}^+)$ three-body state.
The $NN$ channels have, of course, strangeness 0, the $N\Lambda$
channels have strangeness $-1$, and the $\Lambda\Lambda$ and $N\Xi$
channels have both strangeness $-2$. As can be seen from this table,
the $\Lambda\Lambda N$ and $\Xi NN$ systems are coupled together
through the $(i,j)=(0,0)$ two-body channel. 

In Ref.~\cite{Gar07b} we calculated the hyperon-deuteron ($Yd$) scattering lengths as
well as the hypertriton binding energy taking into account all
the $S-$ and $D-$wave components that contribute in the various 
three-body channels. From a combined analysis of the nucleon-hyperon ($NY$) two-body
data, the $Yd$ scattering lengths, and the hypertriton binding
energy, we were able to constrain the allowed values of the 
$\Lambda N$ spin-triplet and spin-singlet scattering lengths 
as $1.41\le a^{\Lambda N}_{1/2,1}\le 1.58$ fm, and $2.33\le a^{\Lambda N}_{1/2,0}\le 2.48$
fm. Therefore, we now make use of the $NY$ interacting models 
satisfying these constraints to calculate the binding energy of
the strangeness $-2$ hypertriton. The results obtained in the
full coupled channel problem $\Lambda\Lambda N$--$\Xi NN$ are show in Table~\ref{t2}.
\begin{table}[h!] 
\caption{Binding energy of the strangeness $-2$ hypertriton (in MeV) 
measured with respect to the $NH$ threshold
for several models of the $NY$ interaction
satisfying the constraints of Ref.~\cite{Gar07b}
for the $N\Lambda$ scattering lengths, $a^{N\Lambda}_{i,j}$ (in fm).}
\begin{ruledtabular} 
\begin{tabular}{cc|cccccc}
 & & \multicolumn{4}{c}{$a^{N\Lambda}_{1/2,1}$} & \\ 
 & & 1.41 & 1.46 & 1.52 & 1.58 \\
\hline
\multirow{3}{*}{$a^{N\Lambda}_{1/2,0}$} & 2.33 & 0.416 & 0.455 & 0.495 & 0.542 \\
& 2.39 & 0.424 & 0.463 & 0.504 & 0.551 \\
& 2.48 & 0.447 & 0.487 & 0.528 & 0.577 \\
\end{tabular}
\end{ruledtabular} 
\label{t2}
\end{table}
As one can see from this table, the strangeness $-2$ three-baryon 
system with quantum numbers $(I,J^P)=(\frac{1}{2},\frac{1}{2}^+)$
is bound, the binding energy varying between 0.4 and 0.6 MeV. 
However, as predicted in Ref.~\cite{Tan65} due to the non-existence
of an isospin one ${}_{\Lambda}^{3}$H bound state, the $\Lambda \Lambda N$ system
alone is not bound. The bound state only appears when the 
coupling between the $\Lambda\Lambda N$ and $\Xi NN$ components
is considered, i.e., when the $(i,j)=(0,0)$ two-body
$t^{\Lambda\Lambda,N\Xi}$ amplitude is included in the calculation.

The relevance of the $\Lambda\Lambda$--$\Xi N$ coupling for double-$\Lambda$ 
hypernuclei has been emphasized for the case of the ${}_{\Lambda\Lambda}^4$H
hypernucleus~\cite{Nem03,Him06}. If this system is studied with $NN$, $N\Lambda$ and $\Lambda\Lambda$
interactions improved for the description of the ${}_{\Lambda\Lambda}^6$He, it 
is found to be unbound. In the case of the ${}_{\Lambda\Lambda}^6$He the 
$\Lambda\Lambda$--$N\Xi$ coupling plays a minor role, because the nucleon
generated in the transition must occupy an excited $p-$shell, the lowest 
$s-$shell being forbidden by the Pauli principle. This is not the case
of the ${}_{\Lambda\Lambda}^4$H, where the nucleon generated by the  
$\Lambda\Lambda$--$N\Xi$ transition can occupy a hole in the lowest
$s-$shell. This effect generates theoretical binding for the 
${}_{\Lambda\Lambda}^4$H~\cite{Nem03} and it is also the responsible
for generating binding in the strangeness $-2$ three-baryon 
system with quantum numbers $(I,J^P)=(\frac{1}{2},\frac{1}{2}^+)$.
It is therefore important to obtain experimental information about
the strength of the $\Lambda\Lambda$--$\Xi N$ coupling. It could be derived 
from the measurement of the ${}_{\Lambda\Lambda}^4$H binding energy.
In the meantime, the only available experimental data is the inelastic
cross section $\Xi^-p \to \Lambda\Lambda$, correctly described 
by the present model (see Fig. 2 of Ref.~\cite{Val10}).

The possible existence of a strangeness $-2$ hypertriton will give 
a strong impact on forthcoming experimental projects as well as 
on-going theoretical studies. Experimentally, it could be
measured in the J-PARC-E07 experiment, where more than $10^3$
$\Lambda\Lambda$--nuclei are expected to be detected by means of    
$\Xi$--capture reactions using different
target nuclei: $C$, $N$, and $O$~\cite{Nak12}.
Theoretically, Lattice QCD has evolved to the point where
the calculation of the binding energy of light nuclei
and hypernuclei with $A\le 4$ and $\hat S \le 2$,
at unphysically heavy light-quark masses, is possible~\cite{Bea12}. Extrapolations to the 
physical light-quark masses have not been attempted because the
quark mass dependences of the energy levels in the light nuclei are not
knwon. Future calculations at smaller lattice spacings and at lighter
quark masses will facilitate such extrapolations and, therefore, comparison
with experiment and, thus, the analysis of the strangeness $-2$ hypertriton. 

In summary, we have solved for the first time the Faddeev equations 
for the bound state problem of the coupled 
$\Lambda\Lambda N$--$\Xi NN$ system to study whether
an hypertriton with strangeness $-2$ may exist or not. 
We make use of the interactions obtained from a chiral quark model
describing the low-energy observables of the two-baryon
systems with strangeness $0$, $-1$, and $-2$ and three-baryon systems
with strangeness $0$ and $-1$. 
The $\Lambda\Lambda N$ system alone is unbound in agreement with the non-existence
of an isospin one ${}_{\Lambda}^{3}$H bound state. However, when the full coupling to
$\Xi NN$ is considered through the $(i,j)=(0,0)$ two-body
$t^{\Lambda\Lambda,N\Xi}$ amplitude, the strangeness $-2$ three-baryon 
system with quantum numbers $(I,J^P)=(\frac{1}{2},\frac{1}{2}^+)$ becomes
bound, with a binding energy of about 0.5 MeV. 

\begin{acknowledgments} 
The authors thank to Dr. K. Nakazawa, Dr. B. Curto, and Dr. V. Moreno for 
enlightening discussions. This work has been partially funded 
by COFAA-IPN (M\'exico), by Ministerio de
Educaci\'on y Ciencia and EU FEDER under Contract No. FPA2010--21750--C02--02 and by the
Spanish Consolider--Ingenio 2010 Program CPAN (CSD2007--00042).
\end{acknowledgments}

\end{document}